\documentclass[conference]{IEEEtran}
\IEEEoverridecommandlockouts

\usepackage{cite}
\usepackage{amsmath,amssymb,amsfonts}
\usepackage{algorithmic}
\usepackage{graphicx}
\usepackage{textcomp}
\usepackage{xcolor}
\usepackage{url}
\def\BibTeX{{\rm B\kern-.05em{\sc i\kern-.025em b}\kern-.08em
    T\kern-.1667em\lower.7ex\hbox{E}\kern-.125emX}}
\begin{document}

\title{What Affects the Performance of Fake Audio Detection? Analyzing Factors in a Continual Learning Setting
}

\author{\IEEEauthorblockN{Yixuan Xiao}
\IEEEauthorblockA{\textit{Institute for Natural Language Processing} \\
\textit{University of Stuttgart}\\
Stuttgart, Germany \\
yixuan.xiao@ims.uni-stuttgart.de}
\and
\IEEEauthorblockN{Ngoc Thang Vu}
\IEEEauthorblockA{\textit{Institute for Natural Language Processing} \\
\textit{University of Stuttgart}\\
Stuttgart, Germany \\
thang.vu@ims.uni-stuttgart.de}
}

\maketitle

\begin{abstract}
The increasing sophistication of deepfake audio generation technologies makes it important to develop robust fake audio detection systems that can adapt over time. This study examines how various factors impact the performance of detection systems in a continual learning setting. We focus on factors such as attacker architectures, attackers' training datasets, speaker diversity, and task order. We evaluate the performance of three detection models trained with four different strategies, including direct fine-tuning, one-class classification, random replay, and Learning without Forgetting. Results show that artifacts from the fake audios might arise from the attackers' training datasets, and simply changing attacker architectures does not sufficiently challenge detection systems. Moreover, task order and speaker diversity can significantly influence performance, with varying degrees of sensitivity across different detection models and training strategies. These insights underline the need for careful consideration of these factors when developing robust detection systems.
\end{abstract}


\section{Introduction}
\label{sec:intro}

Advanced Text-to-Speech (TTS) and Voice Conversion (VC) technologies can produce realistic audio for beneficial purposes but also pose risks like fake news or fraud, thus making Fake Audio Detection (FAD) an important filed in voice privacy and security. As a result, several competitions, including the ASVSpoof \cite{wuASVspoof2015First2015, kinnunenASVspoof2017Challenge2017,todiscoASVspoof2019Future2019,yamagishiASVspoof2021Accelerating2021} and ADD challenges \cite{yiADD2022First2022a,yiADD2023Second}, have been proposed to advance research in the field.
However, most research addressing these challenges focuses on improving FAD performance on static datasets \cite{yi_audio_2023}, which may not reflect the dynamic nature of real-world TTS and VC technologies. This highlights the need for adaptive methods in a \textbf{continual learning} framework, where models are continuously updated to handle new ``tasks\cite{lomonaco_rish_cl_tutorial_2021}." A ``task" typically refers to a specific FAD dataset or subset in recent continual learning FAD studies \cite{maContinualLearningFake2021, zhangYouRememberOvercoming2023b, zhangWhatRememberSelfAdaptive2024a}. These studies primarily aim to improve task performance, with limited analysis of the factors affecting it. On the other hand, it is difficult to analyze because many factors are involved in one task, e.g., multiple attackers, different speakers, and unknown attacker's training data in some datasets \cite{wang2020asvspoof, yiADD2022First2022a}. When performance improves on a task, it is hard to pinpoint which factors caused the improvement.

Typically, only the variety of \textbf{attacker architectures} is considered when creating challenging FAD datasets\cite{frankWaveFakeDataSet2021,zhangOneclassLearningSynthetic2021, zhang_towards_2024}, so the \textbf{attacker's training data} is not mentioned in some datasets. Since attackers are optimized to capture the characteristics of their training data, we argue that different architectures trained on the same dataset may share common traits in the fake audios. These traits could serve as cues for FAD across different architectures. Thus, we suspect various architectures trained on the same dataset does not really build a challenging FAD dataset.

\textbf{Task order} is an important factor in continual learning, with many studies in the field exploring its impact\cite{de2021continual}. However, there has been limited research on its effects in the FAD community \cite{linTheoryForgettingGeneralization2023,leeContinualLearningTeacherStudent2021}.

Finally, we consider \textbf{speaker diversity} to be an important yet less discussed factor. Some studies assume that, despite diversity, the data distribution of genuine audios should be more consistent than that of fake audios. Therefore, to build robust detection models, some works design the loss function to focus on genuine audio\cite{zhangOneclassLearningSynthetic2021}, while others try to maintain the model's performance on genuine audio in continual learning settings\cite{zhangYouRememberOvercoming2023b}. However, with significantly increased speaker diversity ---given the variation in voices and speaking styles---does this assumption still hold true?

To study these factors, we need precise control over them, requiring a more detailed task definition that enables control experiments. As a result, rather than defining a task by an FAD dataset with many attackers, we define it by focusing on a single attacker with a known architecture and training data, along with genuine audios with a known level of diversity. To generalize the findings, experiments are conducted with more than one detection method and training strategy. Three detection models—Light CNN\cite{wuLightCNNDeep2018}, ResNet\cite{resnetFAD}, and wav2vec2AASIT\cite{tak22_odyssey}—and four training strategies—fine-tuning, one-class classification\cite{zhangOneclassLearningSynthetic2021}, random replay, and Learning without Forgetting (LwF)\cite{liLearningForgetting2018}—are used.

To fill the research gap, our \textbf{research questions} are:

\textbf{RQ1} Which aspect of the attacker has greater impact to the detection model's performance, its training data or architecture?

\textbf{RQ2} Do different task orders have the same impact on different detection models and training strategies?

\textbf{RQ3} Do different speaker diversity levels have the same impact on different detection models and training strategies? 

Our experiments\footnote{https://github.com/XIAOYixuan/tomatoDD/tree/icassp25-cl-factors} show that attacker's training data, often overlooked, has an impact as significant as model architecture. Task order affects detection models and strategies differently. For example, LwF is sensitive to task order in smaller models, while One-class Softmax (a one-class learning method), is task-sensitive in larger models, regardless of task order. Speaker diversity generally has no effect, except for One-class Softmax. 

\section{Related Work}

One pioneering work enhances Continual Learning FAD's generalizability by augmenting the classification loss with both LwF and Positive Sample Alignment loss\cite{maContinualLearningFake2021}. The former retains performance on old tasks, while the latter constrains the genuine feature space across tasks. This study primarily evaluates its methods on two types of ASVSpoof19 tasks\cite{wang2020asvspoof}: (1) Task LA, where fake audios are generated by TTS or VC systems, and (2) Task PA, where fake audios involve different acoustic conditions or simple methods like audio replay. We focus on settings similar to Task LA, as it better reflects real-world scenarios where users increasingly use TTS or VC to generate fake audios.
The study also explores four different TTS/VC architectures to create various tasks, but reports only overall performance after all tasks are completed, making it difficult to deeply analyze the factors influencing performance.

Two other related studies, \cite{zhangYouRememberOvercoming2023b} and \cite{zhangWhatRememberSelfAdaptive2024a}, share similar ideas: during weight updates, they modify gradient directions based on the data type. For instance, for genuine audio, the gradient direction should align closely with previous ones. Their work treats an entire FAD dataset, like ASVSpoof19 \cite{todiscoASVspoof2019Future2019} or In-the-Wild \cite{mullerDoesAudioDeepfake2022b}, as a single task.
We choose not to use this task setup because each task can have many different attack types and acoustic environments, complicating the design of controlled experiments to assess each factor's impact. 

Finally, all of the above-mentioned methods typically handle no more than four tasks. Considering the rapid development of TTS and VC methods, we believe it is important to study the performance on a larger number of tasks. Therefore, in our study, we use MLAAD \cite{mullerMLAADMultiLanguageAudio2024} and LibriSpeech \cite{panayotovLibrispeechASRCorpus2015} to form 19 tasks.

\section{Methods}
\label{sec:method}
To approximate real-life scenarios, we begin with a base detection model trained on an initial task. Then we adapt this base model to a sequence of new tasks.
A straightforward training strategy would be to directly fine-tune the base model on new tasks. We will introduce three other training strategies which are considered to be more robust in the continual learning setting.

\subsection{Detection Model}

For detection models, we choose Light-CNN (LCNN) \cite{wuLightCNNDeep2018}, ResNet\cite{heDeepResidualLearning2016}, and wav2vec2AASIST\cite{w2v2FAD}.
These three architectures range from lightweight to heavyweight and have proven to show good performance in FAD tasks. In order to generalize the findings by conducting experiments with different capacities, we specifically choose these models with different sizes as we suggest that model size may serve as a proxy for model capacity.

\noindent\textbf{LCNN} Input: LFCC-delta-double delta. The model has nine convolution layers along with nine Max-Feature-Map (MFM) operations. The MFM operation combines multiple feature maps and returns the element-wise maximum ones to suppress the number of activated neurons. 

\noindent\textbf{ResNet} Input: LFCC-delta-double delta. The model is also a CNN-based one. It uses residual connections to address the vanishing gradient problem, making it easier to train deeper networks. 

\noindent\textbf{wav2vec2AASIST} Input: raw audios. The model uses wav2vec2 to extract features, and feed them into AASIST \cite{jungAASISTAudioAntiSpoofing2021}, which is a graph neural network that model artefacts spanning temporal and spectral domains using attention mechanisms. 

\subsection{Training Strategies}

Apart from direct fine-tuning \textbf{FT}, we also consider three other training strategies: Random Replay, LwF, and OCS (a one-class continual learning method).
We specifically choose these three strategies to cover two real-life scenarios: The first one involves systems that have some access to previous data and can handle the memory burden; Random Replay is well-suited for this situation. The second one involves systems that have no access to previous data (e.g., due to privacy concerns); LwF and OCS are more appropriate.

\noindent\textbf{Random Replay (RE)} Inspired by \cite{pellegriniLatentReplayRealTime2020a}, this method we design keeps a fixed number of past samples in a buffer. After each task, it reduces the number of old samples and adds randomly selected new samples, ensuring an even distribution of samples across tasks.

\noindent \textbf{Learning without Forgetting (LwF)} An LwF loss is added to the binary classification loss to retain learned knowledge by maintaining similar output behaviors between the model trained on the current task and a cached old model trained on previous tasks. KL-divergence is used to measure the difference between the output probabilities of two models.

\noindent \textbf{One-class Softmax (OCS)} This method has shown strong generalizability to new tasks. \cite{zhangOneclassLearningSynthetic2021} assume that genuine audios share more similar characteristics than fake audios. Buiding on this, they propose a new loss which learns an anchor vector \( w \) in the embedded space, aims to draw genuine audios closer to \( w \) with a smaller margin, and push fake audios further away with a larger margin. This work use cosine similarity to measure the similarity between \( w \) and an audio embedding. 

\section{Datasets}

We use ASVSpoof 2019 LA\cite{wang2020asvspoof} to train our base model. MLAAD\cite{mullerMLAADMultiLanguageAudio2024} and LibriSpeech\cite{panayotovLibrispeechASRCorpus2015} are used to construct continual learning tasks. The process is as follows.

\noindent \textbf{MLAAD}
We choose this specific dataset because both the attackers' architectures and training data information are provided (except for griffin\_lim, as it is a phase reconstruction algorithm and does not require training). Also, several attackers in MLAAD are trained on the same dataset, and some attackers share similar architectures, allowing us to study the impact by controlling factors. 
We use the EN subset for fake audios and the US-EN subset for genuine audios in the experiments. There are 19 attackers, each with 1,000 fake samples. We construct 19 tasks, each focused on audios generated by a specific attacker, as shown in Table \ref{tab:tasks}. There are 46,294 genuine audio samples from three speakers, evenly distributed across 19 tasks during training. The dataset for each task is split into training and test sets with an 80/20 ratio.

\begin{table}[ht]
\centering
\begin{tabular}{|c|l|l|}
\hline
\textbf{Task ID} & \textbf{Training Data} & \textbf{Architecture} \\ \hline
T1  &   None      & griffin\_lim            \\ \hline
T2  & Blizzard2013 & capacitron-t2-c50      \\ \hline
T3  & Blizzard2013 & capacitron-t2-c150\_v2 \\ \hline
T4  & Sam          & tacotron-DDC           \\ \hline
T5  & Jenny        & Jenny                  \\ \hline
T6  & EK1          & tacotron2              \\ \hline
T7  & LJSpeech     & fast\_pitch            \\ \hline
T8  & LJSpeech     & overflow               \\ \hline
T9  & LJSpeech     & glow-tts               \\ \hline
T10 & LJSpeech     & vits--neon             \\ \hline
T11 & LJSpeech     & neural\_hmm            \\ \hline
T12 & LJSpeech     & speedy-speech          \\ \hline
T13 & LJSpeech     & tacotron2-DCA          \\ \hline
T14 & LJSpeech     & tacotron2-DDC          \\ \hline
T15 & LJSpeech     & tacotron2-DDC\_ph      \\ \hline
T16 & LJSpeech     & vits                   \\ \hline
T17 & suno         & bark                   \\ \hline
T18 & multi-dataset & tortoise-v2           \\ \hline
T19 & facebook-mms    & tts-eng            \\ \hline
\end{tabular}
\caption{Task ID,  Attacker's Training Data and architecture}
\label{tab:tasks}
\end{table}

\noindent\textbf{LibriSpeech} We use LibriSpeech to improve the speaker diversity level, choosing it specifically because it shares the same acoustic domain (AudioBook) as the genuine audios from MLAAD. We create another set of 19 tasks by replacing the genuine audios with randomly sampled LibriSpeech audios. 
In total, the LibriSpeech subset contains 46,294 audios and 2,339 unique speakers. On average, each task has more than 100 unique speakers. 

\section{Experiments and Results}
\label{sec:pagestyle}

\subsection{RQ 1 Attacker's Training Data vs. Attacker's Architecture}

To avoid potential biases introduced by the task order and specific speaker traits, we randomly shuffled the task orders and used genuine audios from LibriSpeech for each task. The task order is shown below. 

\noindent\textbf{Order 1 (Random)}:
\textbf{T7} $\rightarrow$ \underline{T4} $\rightarrow$ \textbf{T8} $\rightarrow$ T1 $\rightarrow$ T3 $\rightarrow$ \underline{\textbf{T13}} $\rightarrow$ \textbf{T9} $\rightarrow$ T17 $\rightarrow$ \textbf{T10} $\rightarrow$ T2 $\rightarrow$ \underline{T14} $\rightarrow$ T5 $\rightarrow$ \textbf{T11} $\rightarrow$ \underline{T6} $\rightarrow$ \textbf{T12} $\rightarrow$ \underline{T15} $\rightarrow$ T19 $\rightarrow$ T16 $\rightarrow$ T18

\begin{figure}[ht]
    \centering
    \includegraphics[width=0.49\textwidth]{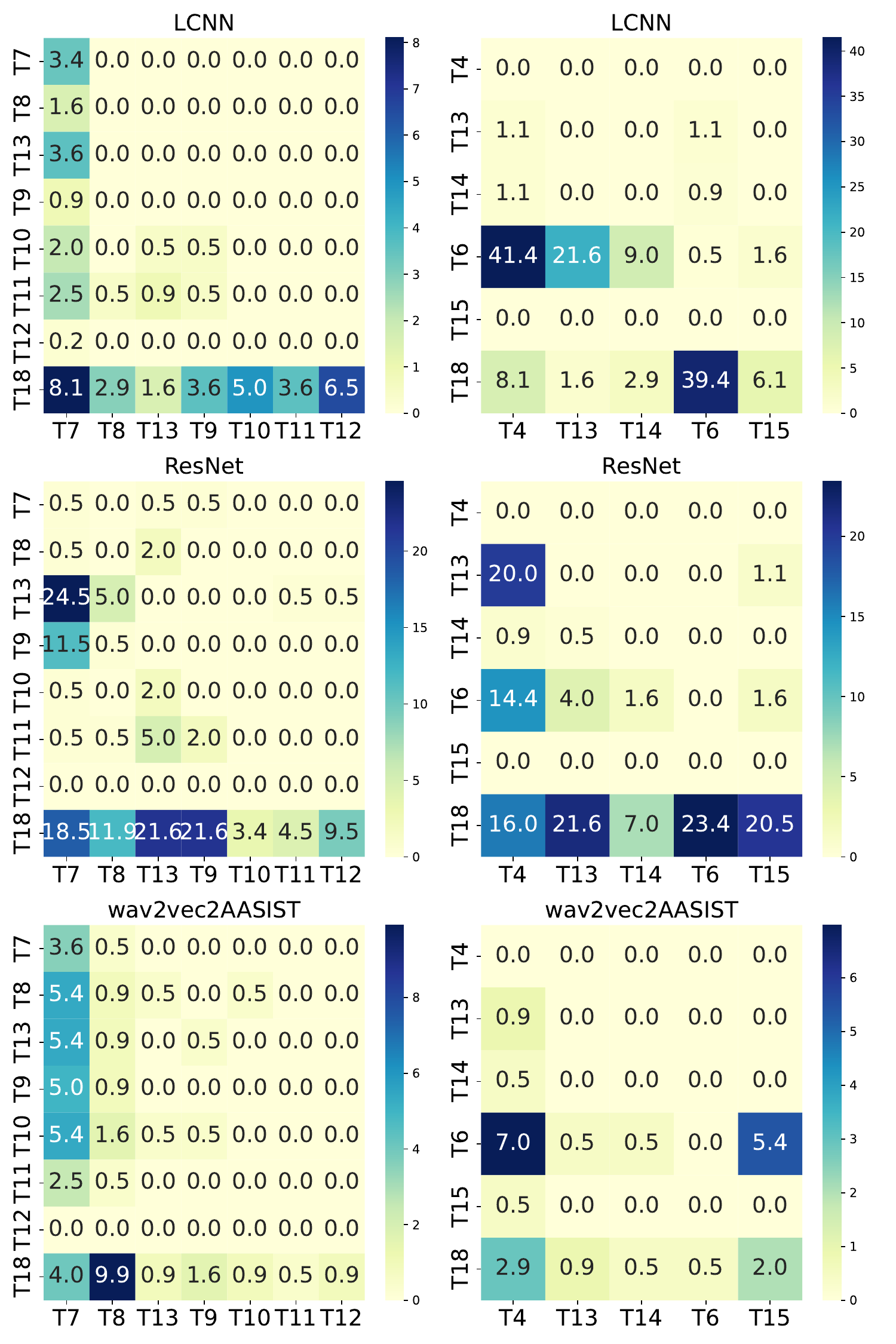}
    \caption{Equal Error Rates (EERs) of Subset 1 and Subset 2 are shown in left column and right column respectively. Lower the better. The x-axis shows the training order.}
    \label{fig:ljspeech_heatmap}
\end{figure}

To analyze the impact of one factor, we need to keep the other fixed. Therefore, results are reported for two subsets: (1) Subset 1: (Tasks are in bold) Attackers have different architectures but trained on the same dataset. (2) Subset 2: (Tasks are underlined), Attackers are Tacotron-based but trained on different datasets. To determine if the impact is specific to each subset or applies to other tasks, we compare with T18, which uses different datasets (mostly online podcasts and audiobooks) and architectures (DALL-E-inspired \cite{ramesh2021zero}). The results of FT training strategy are shown in Figure~\ref{fig:ljspeech_heatmap}.

It is expected that, in Subset 2, EERs on unseen tasks except for T18 decrease once models have seen fake audios generated by attackers with similar architectures. In Subset 1, the improvement on unseen tasks is even stronger. In most cases, models need no more than two tasks to achieve a close-to-zero EER and can maintain low EERs for seen tasks afterwards; while in Subset 2, T6 poses some challenges to all models. 
These results suggest that simply increasing the types of attacker's 
architectures is not enough to create a challenging datasets. Also, results on T6 suggest models seem to be more sensitive to artifacts by varying attacker's training data.

Another observation is that wav2vec2AASIST is the most robust model as it also achieves low EER on T18.
It is likely that self-supervised learning features are more robust than simple acoustic features.

\begin{figure}[ht]
    \centering
    \includegraphics[width=0.5\textwidth]{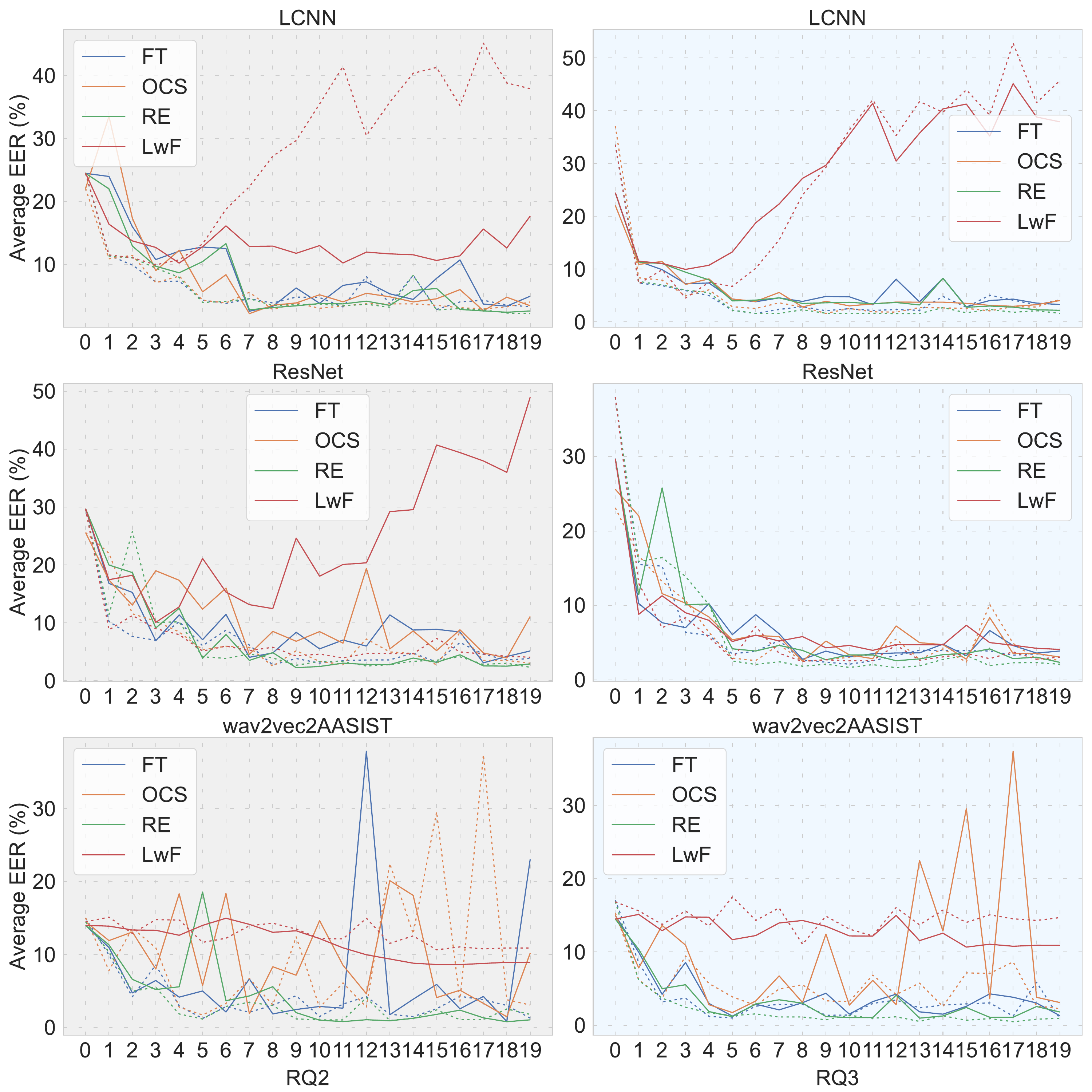}
    \caption{Results for RQ2 and RQ3 are shown in the left and right columns, respectively. In column RQ2, \textbf{dashed} lines are for \textbf{Order1} and \textit{solid} lines are  for \textit{Order2}. In column RQ3, \textbf{dashed} lines are for \textbf{MLAAD} and \textit{solid} lines are for \textit{LibriSpeech}.
}
    \label{fig:rq23}
\end{figure}

\subsection{RQ 2 Impact of Task Order}

Based on previous findings from Subset 1's results, we organize the tasks based on the similarity of the attacker's training data. This setup allows for smoother transitions in data distribution among tasks with similar attacker's training data, but results in more significant shifts between tasks with different ones. The new order is as follow:

\noindent\textbf{Order 2 (By-Dataset)}: 
T1 $\rightarrow$ T2 $\rightarrow$ T3 $\rightarrow$ T4 $\rightarrow$ T5 $\rightarrow$ T6 $\rightarrow$ T7 $\rightarrow$ T8 $\rightarrow$ T9 $\rightarrow$ T10 $\rightarrow$ T11 $\rightarrow$ T12 $\rightarrow$ T13 $\rightarrow$ T14 $\rightarrow$ T15 $\rightarrow$ T16 $\rightarrow$ T17 $\rightarrow$ T18 $\rightarrow$ T19

The results on MLAAD are shown in the left column of Figure~\ref{fig:rq23}.
Among them, RE generally outperforms other strategies for all models, showing the lowest average EERs most consistently without sudden performance drops, which highlights its robustness irrespective of task order.

A notable observation is that LwF can greatly worsen the performance of smaller models such as LCNN and ResNet by changing the task order. For LCNN, Order 2 leads to higher average EERs, while for ResNet, it is Order 1. However, in wav2vec2AASIST, LwF shows a nearly monotonic decrease in average EER from T7 to T16, where the attacker's training data remains the same.
Since LwF aims to approximate the previous model, a smoother transition between consecutive tasks in Order 2 helps the model maintain stable performance. Finally, the different behaviors across models suggest that models with smaller capacities might rely more on shortcuts than general information. Since each task can have unique shortcuts, LwF’s attempt to match the old model's performance can cause unexpected outcomes.

OCS is robust with LCNN, but as model capacity increases, its performance becomes less stable (e.g., average EER fluctuates significantly for wav2vec2AASIST). This shows that while OCS is typically seen as a solution for robust FADs, it can be sensitive to tasks when used with larger models. \textit{At this moment}, we can suggest that the instability may stem from larger embedding dimensions or the model’s ability to capture richer information from genuine audios, complicating the training of a single anchor vector.

\subsection{RQ 3 Impact of Speaker Diversity Level}
This experiment categorizes tasks into two groups based on the datasets the genuine audios originated from: (1) Exp-LibriSpeech: high diversity, and (2) Exp-MLAAD: low diversity. All experiments were conducted using Order 1.
The results are shown in the right column of Figure~\ref{fig:rq23}. For FT, RE, and LwF, performance is similar in both settings. Since speaker information is used to generate fake audios (i.e., similar speaker information is present in both genuine and fake audios), recognizing speaker information does not help much in identifying fake audios. Therefore, increasing diversity should not have a significant impact.

The exception is OCS. Reducing the diversity from high to low stabilizes performance. Combined with the findings from RQ2, it appears that the complexity of genuine audios captured by large models, rather than model capacity itself, makes OCS unstable. We suggest that higher speaker diversity challenges the premise that genuine audios have common characteristics and should share a compact space. In real-life scenarios, with variations such as different recording devices, accents, and background noise, this premise may be challenged more. Further investigation is needed in this direction.

\section{Conclusion}
\label{sec:conclusion}

We study four factors — attacker's training data, attacker's architecture, task order, and speaker diversity — that may affect FAD performance in a continual learning setting. We discover novel findings: \textbf{RQ1} The impact of the attacker's training data is typically overlooked, but it actually has at least as much impact as the architecture. \textbf{RQ2} LwF with smaller-capacity models is sensitive to task order, while OCS, which is typically considered a robust FAD method, is very sensitive to tasks regardless of task order when used with larger models. \textbf{RQ3} Speaker diversity generally does not affect most methods, except for OCS. It is likely that high speaker diversity complicates the genuine audio set, challenging its premise that these audios are similar and should occupy a more compact space.

\bibliographystyle{IEEEtran}
\bibliography{refs}

\end{document}